\documentclass[]{aastex}
\usepackage[onecolumn]{emulateapj5}
\usepackage{graphicx}
\begin{document}
\newcommand{\gmet}{{\rm \bf g}}
\newcommand{\tmet}{\rm \bf T}
\newcommand{\nablamet}{\bf \nabla}
\newcommand{\sonicsub}{\bf sonic}
\title{On Steady State 
Neutrino Heated Ultra-Relativistic Winds from Compact Objects}
\def\question#1{{{\marginpar{\tiny \sc }}}}

\author{Jason Pruet\altaffilmark{1}, George M. Fuller\altaffilmark{1}, and 
Christian Y. Cardall\altaffilmark{2,3,4}}
\altaffiltext{1}{Department of Physics, University of California, San Diego, La
Jolla, California, 92093-0319}
\altaffiltext{2}{Physics Division, Oak Ridge National Laboratory, Oak Ridge,
 	TN 37831-6354}
\altaffiltext{3}{Department of Physics and Astronomy, University of Tennessee,
	Knoxville, TN 37996-1200}
\altaffiltext{4}{Joint Institute for Heavy Ion Research, Oak Ridge National
	Laboratory, Oak Ridge, TN 37831-6374}
\begin{abstract}

We study  steady state winds from 
compact objects in the regime where the wind velocity at 
infinity is ultra-relativistic. This may have relevance to some models of 
Gamma-Ray-Bursts (GRB's). Particular attention is paid to the
case where neutrinos provide the heating. 
Unless the neutrino luminosity is 
very large, $L>10^{54} {\rm erg/s}$, the only allowed steady state solutions
are those where energy deposition is 
dominated by neutrino-antineutrino annihilation at the sonic point.
In this case, the matter temperature near the
neutron star surface is low, less than 1MeV for typical neutrino 
luminosities. This is in contrast to the case for
sub-relativistic winds discussed in the context of supernovae 
where the matter temperature near the neutron
star approximates the temperature characterizing the neutrinos. 
We also investigate the setting of the neutron to proton ratio (n/p) in
these winds and find that only for large ($>10{\rm  MeV}$) 
electron neutrino or electron anti-neutrino temperatures 
is $n/p$ entirely determined by neutrino capture. Otherwise,
$n/p$ retains an imprint of conditions in the neutron star.

\end{abstract}


\section{Introduction} 
\question{added affiliations (CYC)}
\question{Commented out email addresses because they were appearing
under the authors (CYC)}

We examine properties of ultrarelativistic winds from
compact objects and we thereby obtain insights into the special
case where these winds arise from neutrino heating. 
 Our principal motivation for the present study is to 
determine if physically plausible steady state solutions exist in the 
ultrarelativistic 
regime and to determine the relation between neutrino parameters and the
character of the outflow.  A peculiar feature of the steady state ultrarelativistic solutions
is that the matter temperature near the neutron star surface is cold. We comment below
on whether or not this is physically realizable. We also study how the final neutron to 
proton ratio (or alternatively electron fraction, $Y_e=1/(n/p+1)$), is set in these 
models. Only for large ($>10{\rm MeV}$) 
electron neutrino or electron anti-neutrino temperatures does $Y_e$ in the initial 
wind material come into steady state equilibrium with the neutrino fluxes.

There is a rich literature concerning steady state outflow from neutron 
stars. In connection with supernovae the neutrino-driven wind occuring
several seconds post core bounce is interesting because it is the favored 
candidate for the site of the r-process elements \citep{meyer92,wilsonandmathews,takahashi94}. 
The relation between 
neutrino parameters and the wind have been extensively discussed \citep{DSW,QW}.
These studies were concerned with gentle, modest entropy winds, characterized
by Lorentz factors 
of order unity far from the star.
\question{phrase rearranged (CYC)} 
Near the 
neutron star surface these gentle winds are well approximated as 
\question{random text deleted (CYC)}
being in hydrostatic equilibrium. The temperature is set by the competition
between neutrino heating and neutrino cooling terms, i.e., the matter
temperature is roughly equal to the temperature characterizing the
neutrinos. 
Apart from
\question{word changed (CYC)}
effects due to nucleosynthesis, the final electron
fraction in these gentle winds is set by the ratio of electron
neutrino ($\nu_e$) and electron anti-neutrino ($\bar{\nu}_e$) capture rates
on nucleons \citep{Qetal93}. 
The general relativistic extension of the wind equations, redshift factors,
and neutrino trajectory bending have been discussed by \cite{fullerqian96}, 
\cite{Cardall}, and \cite{Salmonson}. 
These works also emphasize 
subrelativistic outflow velocities.
\question{sentence deleted (CYC)}
Detailed studies of relativistic effects along these lines 
for r-process nucleosynthesis have been carried out recently by \cite{japanese1}.
 
In a pioneering study \cite{Paczynski} examined ultrarelativistic flow from neutron stars
relevant for studies of GRB's. In this work no reference was made to a heating
term and, except for photon diffusion which is unimportant near the
neutron star surface, the flow was assumed adiabatic. This work provided 
a detailed picture of the wind evolution far from the neutron star 
surface, which is the regime relevant for optical emission in GRB's. There are many 
similarities between these steady state ultra-relativistic winds and 
the fireballs discussed in connection with GRB's \citep{Piran}.
 
The present study is an attempt to
fill the gap between detailed studies of the relation between neutrino 
heating and sub-relativistic outflow and the study of ultra-relativistic
outflow without reference to a heating mechanism. This is also
a detailed look at the `guts' of a simple model for a GRB central engine.
It may be applicable to the model proposed by \cite{SWM}. 
 In this model binary neutron stars are compressed and heated as 
gravitational radiation saps orbital kinetic energy during the last 
seconds of inspiral. The heating drives copious neutrino loss 
($L_{\nu}\sim 10^{52} {\rm erg \, s^{-1}}$) which in turn drives a
hot pair plasma from the neutron star surface.   

Lastly, it has recently been argued that the electron fraction in neutrino-heated
fireballs may be a clue to neutrino physics at the central engine, and may have
observable consequences \citep{FPA}. The present study serves as a ``proof of 
concept'' of this idea for a simple case. Interestingly, we find that the electron 
fraction in the wind far from the central engine is set by processes very near the
central engine. If this is a generic feature of GRB central engine models,
information about the electron fraction in GRB's may help pin down the central
engine.

\section{Equations}

The general relativistic equations describing the outflow of material from 
the neutron star are Euler's equations: 

\begin{eqnarray} 
{\rm \bf u}\cdot \left(\nablamet \tmet \right) 
&=&- {\rm \bf u}\cdot \left(\nablamet \tmet_{\nu}\right) \cr 
\left(\gmet+{\rm \bf u} \otimes {\rm \bf u}\right)\cdot \left(\nablamet  \tmet \right)
&=&- \left(\gmet+{\rm \bf u} \otimes {\rm \bf u}\right)\cdot \left(\nablamet \tmet_{\nu}\right),
\end{eqnarray}

baryon number conservation,

\begin{equation}
\nablamet \cdot \left(\rho_{\rm b} {\rm \bf u}\right)=0,
\end{equation}

and the change in electron fraction due to lepton capture on baryons,

\begin{equation} 
\label{dye}
{\rm \bf u}\cdot \left(\nablamet Y_e\right)=\left(1-Y_e\right)\left(\lambda_{{\nu}_e{\rm n\rightarrow p}}
 +\lambda_{e^+{\rm n\rightarrow p}}\right)-
Y_e\left(\lambda_{\bar{\nu}_e{\rm p\rightarrow n}}+\lambda_{e^-{\rm p\rightarrow n}}\right).
\end{equation}

In the above ${\rm \bf u}$ is the 4-velocity 
of the outflowing material, $\tmet$ and $\tmet_{\nu}$ are the stress-energy tensors 
for the outflowing matter and the neutrinos respectively, $\gmet$ is the metric 
tensor, and $\rho_{\rm b}$ is the rest mass energy density of baryons in the plasma
comoving frame. In Eq. (\ref{dye}) $\lambda_{\nu_e{\rm n\rightarrow p}}$ and 
$\lambda_{\bar{\nu}_e{\rm p\rightarrow n}}$ are the capture rates for electron anti-neutrinos
and electron neutrinos on neutrons and protons respectively, and 
$\lambda_{e^+{\rm n\rightarrow p}}$ and $\lambda_{e^-{\rm p\rightarrow n}}$ are 
the capture rates for positrons and electrons on neutrons and protons respectively.
In what follows we employ the Schwarzschild metric and use 
natural units,  $G=c=\hbar=k_b=1$, with $k_b$ the Boltzmann constant.
In 
\question{word deleted (CYC)}
Schwarzschild coordinates the line element is  
\begin{equation}
ds^2=\gmet _{\mu \nu} dx^{\mu} dx^{\nu}=
-\left(1-2M/r\right)dt^2+\left(1-2M/r\right)^{-1}dr^2+r^2d\Omega^2, 
\end{equation}
Here $M$ is the mass of the neutron star and $d\Omega^2=d\theta^2+\sin^2{\theta}d\phi^2$. 
Imposing steady state conditions and writing the radial component of the
4-velocity ${\rm u}^r$ as ${\rm  u}^r=vy$ where $y=\left(1-2M/r\right)^{1/2}/\left(1-v^2\right)^{1/2}$ and 
$v$ 
is the outflow velocity of the plasma as measured in an inertial frame
at rest in the Schwarzschild coordinate ($t,r,\theta,\phi$)  system allows us to rewrite 
the equations above as

\begin{mathletters}
\begin{eqnarray}
\label{arhob}
(1-v^2)a \rho_{\rm b}&=&\left(\rho + P\right) \left[{vv' \over 1-v^2} + {M \over r^2}{1 \over \left(1-2M/r\right)}\right]
+P'  \\ \cr
\label{rhob}
q \rho_{\rm b}&=&vy\left[\rho '+\left(\rho + P\right)\left({v'\over v(1-v^2)}+ 
{M \over r^2}{1 \over \left(1-2M/r\right)}+{2\over r}\right)\right] \\ \cr
\label{mdot}
\dot M&=&4\pi r^2\rho_{\rm b}vy  \\ \cr
 vy Y_e ' &=& \left(1-Y_e\right)\left(\lambda_{{\nu}_e{\rm n\rightarrow p}}
 +\lambda_{e^+{\rm n\rightarrow p}}\right)-
Y_e\left(\lambda_{\bar{\nu}_e{\rm p\rightarrow n}}+\lambda_{e^-{\rm p\rightarrow n}}\right).
\end{eqnarray}
\end{mathletters}
In the above, primes denote differentiation with respect to r, and 
$\rho$ and $P$
are the total energy density (including rest mass) and pressure, respectively, 
of the plasma in the
comoving plasma frame. We have also introduced 
$a=- \left(\left(\gmet+{\rm \bf u} \otimes {\rm \bf u}\right)\cdot \left(\nablamet  \tmet_{\nu}\right)\right)_r/\rho_{\rm b}$ and 
$q= -{\rm \bf u} \cdot \left(\nablamet  \tmet_{\nu}\right)/\rho_{\rm b}$ which are respectively the 
momentum and energy deposition into the plasma per unit time per unit baryon
rest mass. $\dot M$ is the mass outflow rate. The second law of 
thermodynamics and some algebra allows us to produce from these expressions two more first integrals of the flow:
\begin{eqnarray}
\label{firstintegral}
\frac{ds}{dr}&=&\frac{{ m_{\rm b}}q}{T vy}\\
\label{secondintegral}
\frac{d (\hat H y)}{dr}&=&y \left({(1-v^2)a+q/(vy)}\right).
\end{eqnarray}

Here we have introduced the nucleon mass ${ m_{\rm b}}$, the 
entropy per baryon $s$ (in units of Boltzmann's constant), and the enthalpy
$\hat H=(\rho+P)/\rho_{\rm b}$ of the plasma. It is useful to have explicit formulas for these
quantities:

\begin{mathletters}
\begin{eqnarray}
\label{ent}
s=5.21 \rho^{-1}_8 \left({T\over {\rm MeV}}\right)^3  \\ \cr
\label{enth}
\hat H -1 \equiv H={Ts \over m_{\rm b}}
\end{eqnarray}
\end{mathletters} 

In this equation $\rho_8$ is the density in units of $10^8{\rm g cm^{-3}}$.
The number 5.21 above
becomes 1.89 as $T$ decreases below the threshold for pair production.
 In writing Eqs. \ref{ent} and \ref{enth} we have
assumed that the electrons are non-degenerate and that the entropy of 
light particles dominates the entropy of nucleons (i.e. $s>20$). This is
relevant for our analysis because we will   
be concerned principally with conditions near the sonic point, and we
will show below that low entropy conditions near the sonic point cannot
lead to ultrarelativistic flow at infinity. The quantity H is related 
to the quantity $\eta$, the ratio of energy density to rest mass energy 
density often discussed in connection with GRB's, through $\eta \sim (3/4)H+1$.

If the flow is to be relativistic at infinity it must either be supersonic
at the neutron star surface or pass through a sonic point somewhere above the neutron star. 
In the present 
work we focus on the case where the flow is not already supersonic at the neutron 
star surface. For the case of nondegenerate electrons and negligible baryon pressure ($s>20$) the 
equation governing $v'$ can be written as

\begin{equation}
{1\over {v\left(1-v^2\right)}}\left(1+H\right)\left(v^2-v_s^2\right)v'={1\over r}\left[{2\over3} H-\left(1+2H/3\right)\left(M/r\right)
{1\over {1-2M/r}}+ ar\left(1-v^2\right)-{qr \over 3{\rm  u}^r}\right]
\end{equation}
From this equation we see that at the sonic point the following conditions
must be satisfied, 

\begin{mathletters}
\begin{equation}    
v^2=v_s^2={H \over 3}{1 \over H+1} 
\end{equation}
 and

\begin{equation}
\label{sonic3}
\left(1-g\left(r\right)\right){2 \over 3} H=g(r)+{qr \over 3{\rm  u}^r}-\left(1-v^2\right)ar.
\end{equation}
\end{mathletters}
In this last equation we have introduced 
\begin{equation}
g\left(r\right)\equiv M/\left(r-2M\right)
\end{equation}
and ${v_s}^2=(dp/d\rho)$ is the sound speed with the differentiation is taken
at constant entropy.

Eq. (\ref{sonic3}) implies that at the sonic point $qr/{\rm  u}^rH<g(r)$. This is interesting
because the change in entropy of the outflowing gas past the sonic point is implicitly  
given by 

\begin{equation}
\label{ineq}
\int{ds\over s}=\int{{ q m_{\rm b} \over Ts}{dr\over {\rm  u}^r}}=\int{qdr\over {\rm  u}^rH}<
\left({q r \over
{\rm  u}^r H}\right)_{\sonicsub} \sim 1,
\end{equation}

Here the subscript $s$ denotes values at the sonic point and the 
second to last inequality holds if $q/{\rm  u}^rH$ decreases more rapidly than $r^{-1}$.
When the inequality in Eq. (\ref{ineq}) holds, then, the entropy is approximately constant past
the sonic point. Because the hydrodynamic equations imply that 
$3T'/T=s'/s-2/r-({{\rm  u}^r} '/{\rm  u}^ r)$, the inequalities in Eq. (\ref{ineq}) also imply that the 
temperature cannot increase substantially past the sonic point for accelerating 
flow. This implies
that the enthalpy per baryon is roughly constant or decreasing past the sonic point. We then arrive
at the important result that that the final 
Lorentz factor of the wind is not 
much larger than the value of $H$ at the sonic point when the above inequality 
is satisfied.  (The final Lorentz factor of the wind occurs when when the energy in relativistic 
particles $\sim H\rho_{\rm b}$ has been converted to baryon kinetic energy.)
This will allow us to make important conclusions regarding 
the sonic point conditions needed in order that the final flow be relativistic.

We can roughly divide the possible sonic point solutions into two families:
those for which the terms involving the heating and momentum deposition terms
q and a may be neglected (Family I),
and those for which these terms are important in determining
the conditions at the sonic point (Family II).  As we will show
that the neutrino energy deposition terms satisfy the inequalities in 
Eq. (\ref{ineq}) when
H and ${\rm u}^r$ increase with radius, only family II solutions can
support steady state ultrarelativistic flow. An exception to this occurs when
the sonic radius is very near 3M.
Previous discussions of 
steady state winds from neutron stars have focused on family I solutions.
For these solutions $H\sim (3/2)M/(r-3M)$ at the sonic point, 
and hence the sonic point must be near 3M
in order to produce ultrarelativistic flow in the absence of heating terms
\citep{Paczynski}. For neutrino driven wind studies relevant for the r-process
the heating terms are generally not important near the sonic point simply
because those flows are subrelativistic or mildly relativistic,
 so that the sonic point occurs far 
from the neutron star where neutrino heating has fallen off.

For Family II solutions (those where heating determines the sonic point conditions
and drives the flow relativistic) some general observations can be made. If we 
assume that $H(r_{s})\gg H(r_0)$ and $s(r_{\sonicsub})\gg s(r_0)$ (where $r_{\sonicsub}$ is the schwarzschild
radial coordinate of the sonic point and $r_0$ is the initial radial coordinate of the flow, usually
taken to be the neutron star radius here) then the sonic
point condition and Eq. \ref{secondintegral} imply

\begin{equation}
\label{yeff}
\int^{r_{\sonicsub}}_{r_0}{\frac{yq}{{\rm u}^r}}\equiv y_{eff}\int^{r_{\sonicsub}}_{r_0}{\frac{q}{{\rm u}^r}}
\sim \left(\frac{yqr}{2{\rm u}^r(1-g)}\right)_{\sonicsub}.
\end{equation}
In Eq. \ref{yeff} we have defined $y_{eff}$, we have assumed that $(1-g)H\gg 1$ at the 
sonic point, and we have neglected the momentum deposition term $a$. 
Eq. \ref{yeff} 
serves to specify the position of the sonic radius given a heating term $q$ 
and a velocity profile. If $q/{\rm u}^r \propto r^{-\beta}$ (which occurs in steady state
winds if the heating rate per unit volume drops as $r^{-\beta-2}$) the sonic point
occurs at

\begin{equation}
\label{rsonicarbitrary}
\frac{r_{\sonicsub}}{r_0} \sim \left(\frac{2(\beta-1)y_{\sonicsub}}
{y_{eff}(1-g_{\sonicsub})}+1\right)^{1/(\beta-1)},
\end{equation}
which implies that the sonic point occurs near $r_0$ for rapidly
dropping heating terms. Again the momentum deposition term $a$ has
been neglected in this approximation.  Another interesting
property of these winds is that the temperature does not vary greatly 
within the sonic radius if $q$ is a decreasing function of radius. This is 
seen by noting that Eqs. \ref{firstintegral} and \ref{yeff} imply

\begin{equation}
\label{teff}
T_{s}^{-1}\sim\left(\frac{y_{s}}{y_{eff}}\right) \frac{\int^{r_{\sonicsub}}_{r_0} {q}/{T {\rm u}^r}}  
 {\int^{r_{\sonicsub}}_{r_0} {q}/{{\rm u}^r}},
\end{equation}
which allows us to to define an effective temperature within the sonic point,
$T_{eff}=T_{s} y_{s}/y_{eff}$.

The above equations allow us to extract the properties of
ultrarelativistic steady state winds for arbitrary heating
functions. We now turn to the particular case where neutrinos from the
central compact object are responsible for the heating.

\section{Neutrino driven ultra-relativistic winds}

Perhaps the first question raised by a study of steady state
neutrino-heated wind solutions is whether or not these
winds are physically realizable for conventional compact objects. 
There are two parts to this question:
i) whether or not steady state can be achieved; and ii) whether or not the 
flow can be relativistic. We defer consideration of the second part until
later, when we discuss mass ablation from the neutron star surface. 
While a hydrodynamic simulation is needed to determine the timescale 
to achieve steady state in these winds, a rough estimate is that  
the equilibration 
timescale is of order the sound crossing time between the sonic radius 
and the neutron star surface, $\sim 10^{-4}{\rm s}$. This timescale
is short compared to the neutrino diffusion timescale, which is of the 
order of seconds. The neutrino diffusion timescale governs the 
evolutionary timescale of the neutron star unless it undergoes a phase
transition or becomes dynamically unstable. It is known that for most cold
neutron star equations of state  general relativistic instability 
sets in for $r\sim 3M$.
 Below this radius the star becomes dynamically
unstable and collapses on a dynamic timescale which is 
comparable to the equilibration timescale
given above. For these small radius stars, then, it may not be sensible to 
put much stock in a 
\question{extra ``a'' deleted (CYC)}
steady state wind solution.

Before describing in some detail properties of the allowed steady 
state solutions, we must first examine the neutrino energy deposition terms.
To parametrize the neutrinos we make the usual assumption that the neutrinos
are emitted from a neutrinosphere with some radius $r_0$ (in this work we
do not make a distinction between neutrinosphere radius and neutron star 
radius) and 
are characterized at the neutrinosphere by a Fermi-Dirac black body 
energy distribution
with temperature $T$ and zero chemical potential,

\begin{equation}
f_{\nu}={{1}\over {e^{-E/T}+1}}.
\end{equation}

This is a crude approximation to the actual expected energy spectra but 
suffices to mock-up the energy deposition physics (see {\cite{Cardall}}).
We further make the free streaming approximation, i.e. we neglect 
the fact that the neutrinos suffer a small depletion with increasing 
radius owing to interactions with the outflowing plasma.
The facts that real neutron stars are characterized by some small but finite
decoupling region over which the neutrino distribution function continues to 
evolve and that neutrinos from real neutron stars are in general non-thermal
or better characterized by degenerate spectra are not important for 
our arguments. For a discussion of when it is apropriate to treat the 
neutrino as sharply decoupling from a neutrinosphere see \cite{Salmonson}.

Evaluating $\nablamet  T_{\nu}$ for various neutrino plasma 
interactions is most easily accomplished by noting that Liouville's
theorem implies that in an inertial frame at rest at some Schwarzschild 
coordinate r the neutrino distribution function is characterized
by the red-shifted temperature 
\begin{equation}
T_{\nu}\left(r\right)=T_{\nu}(r_0)\left({1-2M/r_0\over 1-2M/r}\right)^{1/2}\equiv h T_{\nu}(r_0),
\end{equation}  (see \cite{fullerqian96}) which also
serves to define the redshift factor $h$, and a maximum
angle of deviation from the radial direction given by 
\begin{equation}
\cos \theta_{max}\equiv
 x=\left(1-\left({r_0\over r}\right)^2{1-2M/r \over 1-2M/r_0}\right)^{1/2}.
\end{equation}

 The mass $M$ of the
neutron star appears in the expresion for $x$ because of the bending of 
null trajectories in the Schwarzschild geometry as has been discussed by 
\cite{Cardall} and \cite{Salmonson}.

In an inertial frame at rest in Schwarzschild coordinates, the 
contribution to $\nablamet  T_{\nu} \equiv 
\left(Q^{\hat 0}_{\nu \bar \nu},Q^{\hat r}_{\nu \bar \nu}\right)$
from neutrino-antineutrino annihilation is

\begin{eqnarray}
Q^{\hat 0}_{\nu \bar \nu}&=& {G_{\rm F}^2 \over 9 (2\pi)^5}C T_{\nu}\left(r\right)^4 T_{\bar \nu}\left(r\right)^4F_4\left(0\right)F_3\left(0\right)\bigl(T_{\nu}\left(r\right)+T_{\bar \nu}\left(r\right)\bigr)
\Phi \left(x\right) \cr
Q^{\hat r}_{\nu \bar \nu}&=& {G_{\rm F}^2 \over 9 (2\pi)^5}C T_{\nu}\left(r\right)^4 T_{\bar \nu}\left(r\right)^4F_4\left(0\right)F_3\left(0\right)\bigl(T_{\nu}\left(r\right)+T_{\bar \nu}\left(r\right)\bigr)
 \Psi \left(x\right)/4
\end{eqnarray}

In the above $\Phi(x)=(x-1)^4(x^2+4x+5)$, 
$\Psi(x)=(x-1)^4(x+1)(3x^2+9x+8)$, $G_{\rm F}$ is the Fermi constant, and hats
are used to denote quantities in an inertial frame at rest in the Schwarzschild
coordinate sytem. The quantities $F_3(0)=7\pi^4/120$ and $F_4(0)\approx 23.3$ are the  
Fermi-integrals of argument zero. The quantity $C$ depends on the neutrino species. For
$\nu_e \bar{\nu}_e$ annihilation $C=(1+2 \sin^2 \theta_w)^2+4\sin^4\theta_w$ 
where $\theta_w$ is the Weinberg angle 
($\sin^2\theta_w \approx 0.23$), while for $\nu_{\mu} \bar{\nu}_{\mu}$ and  
$\nu_{\tau} \bar{\nu}_{\tau}$ 
annihilation $C=(1-2 \sin^2 \theta_w)^2+4\sin^4\theta_w$.
For simplicity in what follows we will assume that the luminosity of 
a single neutrino flavor gives all $\nu \bar{\nu}$ annihilation heating
and that for this flavor
the neutrino and antineutrino temperatures are equal, 
$T_{\nu}=T_{\bar{\nu}}$. We will refer to this single 
neutrino flavor as $\mu$ or $\tau$. This is merely a calculational
device designed to reproduce the total heating rate from all neutrino flavors
and adjusted accordingly for sensitivity studies.

Making a transformation back to Schwarzschild coordinates,
we find that 
$\rho_{\rm b} q_{\nu \bar{\nu}}=(1-v^2)^{-1/2}Q^{\hat 0}_{\nu \bar \nu}
(1-v\Psi/4\Phi)$ and 
$\rho_{\rm b} a_{\nu \bar{\nu}}=(1-v^2)^{-1}(1-2M/r)^{-1/2}Q^{\hat 0}_{\nu \bar \nu}
(\Psi/4\Phi-v)$. The terms entering the right hand side 
of the sonic point condition are then

\begin{eqnarray}
{r q_{\nu \bar{\nu}} \over 3{\rm u}^r}&=&{10^{-9} r_6 \over(1-v^2)^{1/2} {\rm u}^r} 
 {\left(T_{\nu}(r)/{\rm MeV}\right)^9\over \left(T/{\rm MeV}\right)^4} \Phi H \left(1-{v\Psi 
\over 4\Phi}\right)\cr
a_{\nu \bar{\nu}} \left(1-v^2\right)r&=& 
{3 \cdot 10^{-8} r_6 \over (1-2M/r)^{1/2}}  
 {\left({T_{\nu}(r)/{\rm MeV}}\right)^9 \over \left(T/{\rm MeV}\right)^4} \Phi H 
\left({\Psi \over 4\Phi}-v\right),
\end{eqnarray}
where $r_6=r/10^6 {\rm cm}$.

Evaluation of the contribution to $(\nablamet  T_{\nu})$ 
from neutrino-plasma interactions is complicated by the fact that an inertial 
observer comoving with the outflowing plasma sees a neutrino distribution
function whose temperature varies with direction as $T_{\nu}(r)
(1-v \cos \theta)/(1-v^2)^{1/2}$. The effect of this velocity (and direction)
dependent redshift is to make the neutrino-plasma heating rates small as
the outflow velocity becomes large.
We will show that neutrino-antineutrino annihilation
is the dominant heating term for the conditions of interest here. It 
suffices for us to note that in the limit where $v=0$, the contributions
to  $(\nablamet  \tmet_{\nu})$ from neutrino-electron scattering and
neurino capture on baryons are, respectively, 

\begin{equation}q_{\nu e} \sim 10 \,q_{\nu \bar{\nu}}  {T^4 \over T_{\nu}^4} {1 \over h^4} 
{1-x \over \Phi} 
\end{equation}

if $T<T_{\nu}$, and 

\begin{equation}
q_{\nu b} \sim 10^3 q_{\nu \bar{\nu}}  
{T^4 \left(T_{\bar{\nu}_e}^6 Y_e +T_{{\nu}_e}^6 \left(1-Y_e\right)\right) \over m_{\rm b} h^3 H T_{\nu}^9}
{1-x \over \Phi}.   
\end{equation}

Note that all of these heating terms satisfy the inequality in 
Eq. (\ref{ineq}) if 
$H$ increases with radius. Neutrino-Antineutrino annihilation 
satisfies Eq. (\ref{ineq}) because $q_{\nu \bar{\nu}}/H \sim \Phi r^2/(\dot M H)$ which falls 
more rapidly than $r^{-6}$. Neutrino electron scattering satisfies 
Eq. (\ref{ineq})
because $q_{\nu e}/H\sim (1-x) \sim r^{-2}$ Neutrino capture 
on baryons also satisfies this condition 
because $q_{\nu b}/H \sim (1-x)/H$. These estimates for the scaling
of the heating rates with radius neglect the dependence of the heating 
on the redshift factor $h$. This implies that the heating rates decrease even
more rapidly with increasing radius if the wind is accelerating. 
This means that if 
the final flow is to be ultra-relativistic, we must have $H>1$ at 
the sonic point, 
i.e., either $r_{\sonicsub}\sim 3M$, or, 

\begin{mathletters}
\begin{equation}
\label{nunubardom}
T\sim 1\, {\rm MeV}\, \left({r_6 \Phi\left(1-g\right) \left( {T_{\nu}(r)/{\rm 10 MeV}}\right)^9 \over {\rm u}^r}\right)^{1/4}
\end{equation}

for $\nu\bar{\nu}$ annihilation dominating the heating, 
or

\begin{equation}
\label{nuedom}
T_{\nu} > 40 \, {\rm MeV} \, h^{-1} \left({1-g \over r_6(1-x)}\right)^{1/5}
\end{equation}

if electron-neutrino scattering dominates the heating rate, or,

\begin{equation}
\label{nubdom}
T_{\nu} > 30 \, {\rm MeV} \, h^{-1}  \left({1-g \over r_6(1-x)}\right)^{1/6}
\end{equation}
\end{mathletters}
if neutrino capture on baryons dominates. In Eqs. (\ref{nunubardom},\ref{nuedom},
and \ref{nubdom}) the vvalues of all quantities 
are taken at the sonic point.

Unless the sonic radius occurs near $3M$ ($g\sim1$), the neutrino 
temperatures needed to drive steady state ultrarelativistic flow 
imply extraordinary neutrino luminosities. (For example, consider the cases inherent in Eqs.
\ref{nuedom} and \ref{nubdom}, 
$L\approx 10^{54} r_6^2 (T_{\nu}/20{\rm MeV})^4 {\rm erg/s}$.) In what follows
we therefore restrict ourselves to the case where $\nu \bar{\nu}$
annihilation dominates the heating at the sonic point and where only 
smaller, more realistic, neutrino luminosities
are required.

\subsection{Steady state flow when $\nu\bar{\nu}$ annihilation dominates at the sonic
point}

Having determined that for modest neutrino luminosities 
($L<10^{54} \rm{ergs \,s^{-1}}$) ultrarelativistic steady state flow 
only occurs if $\nu\bar{\nu}$ annihilation heating 
dominates at the sonic point, we 
now turn to a discussion of the behavior of these solutions.

For $\nu\bar{\nu}$ annihilation the heating scales as  
$q/{\rm u}^r\propto \Phi r^2 h^9 +O(v)$
 and the sonic radius is readily estimated. This is shown in 
Fig. 1. From this figure we see that 
the sonic radius is generically near $r_0$. Because $q_{\nu \bar{\nu}}/{\rm u}^r$ decreases
rapidly with radius we may approximate $y_{eff}\approx y(r_0)$ and $T_{eff}\approx
T(r_0)\approx (y_{\sonicsub}/y_{eff})T_{\sonicsub}$, which implies 
that the temperature is ``cold'' within the sonic radius so long as
thermal 
neutrino energy losses are unimportant. Here ``cold'' means smaller than the neutrino 
energy distribution temperatures.

The mass ablation rate for $\nu \bar{\nu}$ dominated solutions is 
\begin{equation}
\dot M \approx 10^{-4} \left({M_{\odot} s^{-1}} \right) 
\left(\frac{r_6^3 \Phi (1-g) (T_{\nu}(r)/{\rm 10 MeV})^9}{H}\right)_{\sonicsub}.
\end{equation}
If $\dot M$ is too large the flow cannot be relativistic. This is the
``baryon loading problem'' discussed in connection with GRB's. The relation
between $\dot M$ and the neutrino heating rates depends in detail 
on the structure of the compact object atmosphere below, above, and through the neutrinosphere.
As argued by \cite{woos1}, obtaining a sufficiently low $\dot M$ may be particularly difficult 
for $\nu \bar{\nu}$ energy deposition because this heating rate drops 
so rapidly with radius. This requires the density scale height above the
neutrinosphere to be small. For type II supernovae, calculations show that 
$\dot M \approx \dot E_{\nu}/(GM/r_0)$, where $\dot E_{\nu}$ is the 
integrated neutrino energy deposition rate \citep{wilsonandmathews}. 
This implies that most of the neutrino energy deposition 
goes into extracting baryons from the gravitational potential well 
of the neutron star.
  For supernovae, then, ultra-relativistic
winds are not expected unless the early-time density scale height 
is smaller than expected or the late time neutrino luminosities are 
larger than expected. \cite{woos1} 
have argued that the baryon loading problem
cannot be overcome for mass ejection from strange stars resulting
from the phase transition of a cooling or accreting neutron star while 
\cite{SWM} have argued that neutrino heating during the compression of
an inspiralling neutron star might lead to ultrarelativistic flow. 
The question must be considered for each proposed GRB site.

An estimate of $T_{eff}$ allows a simple determination 
of the influence of neutrino
capture on the electron fraction in the outflow. Noting that 
the number of neutrino captures per baryon $n_c$ is 
approximately given by $dn_c=dr \, q_{\nu {\rm b}} /(T_{\nu} {\rm u}^r)$, where
$\nu=\nu_e$ or $\bar{\nu}_e$ for neutrino capture on neutrons or protons, gives
the total number
of neutrino captures per baryon between the neutron star surface and the
sonic point 
\begin{equation}n_c\approx 
{10^{-6}r_6(h^5(1-g_{})(1-x_{}))_{\sonicsub}(T_{\bar{\nu_e}}^5+T_{\nu_e}^5)} 
\ln\left({s_{\sonicsub}/s_0}\right),
\end{equation}
where $s_{\sonicsub}$ and $s_0$ are the entropy at the sonic point and
neutron star surface, respectively. With this expression we can solve for 
the final electron fraction in the fireball, neglecting $e^+/e^-$ capture
because of the low plasma temperature. Representive solutions are shown
in Fig.2. This figure serves to illustrate the basic features of how the 
final electron fraction 
is set: i) For low $T_{\nu_e},T_{\bar{\nu}_e}$, $Y_e$ 
simply remains what it was
in the neutron star. ii) As  $T_{\nu_e},T_{\bar{\nu}_e}$ increase, the final 
$Y_e$ tends
toward equilibrium with respect to neutrino capture (i.e. $Y_e\rightarrow
T_{\nu_e}^5/(T_{\nu_e}^5+T_{\bar{\nu_e}}^5)$), while still retaining
some memory of the value of the electron fraction in the neutron star
crust. And iii) For  $T_{\nu_e}$ or $T_{\bar{\nu}_e}$ $>10{\rm MeV}$, 
$Y_e$ is set
ultimately by the competition between $\nu_e$ and $\bar{\nu_e}$ captures.
 If the neutrino temperatures
are high and neutrinos set the electron fraction in the outflow, 
the final electron fraction is expected to be low because the 
neutrinos emitted from a neutron rich star generically satisfy the 
temperature hierarchy $T_{\nu_e}\le T_{\bar{\nu_e}}$. Because neutron star crusts
typically have $Y_e<0.1$, the electron fraction in the outflow is also 
expected to be low even when neutrino capture does not set the final 
electron fraction.

A numerical example of a steady state neutrino heated wind leading to 
ultrarelativistic flow is shown in Fig 3. For this example we have taken
$M=1.4M_{\odot}$, $r_0=4M$, $T_{\nu_{\mu}}=T_{\bar{\nu}_{\mu}}=10\,{\rm{MeV}}$, and 
$\eta=4/3 H=100$ at the sonic point. As the wind behavior shown in figure 1 is quite
independent of $H$ (except of course for the baryon density which scales trivially with $H$)
as long as $H>$ a few, we have only shown a single numerical example. 
The sonic radius can be chosen to 
match a particular entropy at the neutron star surface. Increasing $r_{\sonicsub}$ increases the
entropy at the surface. The value of the surface entropy is very sensitive to changes in $r_{\sonicsub}$.
Numerically, for $r_{\sonicsub}$ too large, the scale height for velocity changes becomes very 
small and the integration fails. Physically, the sonic radius is set by the requirement that 
the energy flow $\dot E$ at the sonic radius equals the net neutrino energy deposition rate
interior to the sonic radius. Our plot only extends out to 
$r \approx 3r_0$. Past this 
radius neutrino energy deposition is unimportant and the flow satisfies the 
simple scaling laws given in \cite{piran2}. The final Lorentz factor of the wind is approximately 150.

\section{Conclusions}

We have extended the study of steady state winds from compact 
objects to include the case where the wind is driven by a heating
term and the wind velocity at infinity is ultra-relativistic. We find that 
for heating
rates per unit volume which drop more rapidly than $r^{-3}$, the sonic point 
condition is dominated by the heating and momentum deposition terms, 
the sonic radius is near the compact object radius, and the temperature
within the sonic radius is roughly constant. 

Particular attention has been paid to the case where the wind is 
driven principally by neutrino energy deposition. For these winds 
neutrino anti-neutrino annihilation dominates the heating unless the
neutrino luminosity is very large ($\sim 10^{54} {\rm erg s^{-1}}$). Interestingly,
then, for a given neutrino luminosity there are an infinite family of 
steady state supersonic solutions. On one branch of this family is the usual
subrelativistic solution discussed for supernovae, and on the other their
exists
a continuum of ultra-relativistic solutions dominated by neutrino 
heating at the sonic point. Of course, the low mass ablation rate needed
for the ultra-relativistic solutions may not be compatible with the
structure of the atmosphere of the compact object near the neutrinosphere.
This needs to be investigated for each compact object.

The way in which the electron fraction is 
set in the steady state winds we have discussed 
depends sensitively on the neutrino luminosity, with the number of 
neutrino captures per baryon varying as the fifth power of 
the neutrino temperature and depending only logarithmically on the 
final Lorentz factor of the outflow. For high
$\nu_{e}$ or $\bar{\nu}_e$ temperatures the final electron fraction is 
set by neutrino capture. Otherwise $Y_e$ remembers its value in the
neutron star crust.  At least for a simple case, then,
the electron fraction in the outflow is a diagnostic of
conditions within the central engine.
This is interesting because {\it a priori} one might
guess that the electron fraction comes to equilibrium at $Y_e=1/2$, or is 
always dominated by neutrino capture, as is the case for 
supernovae. It is also interesting because recently \citep{bahcall} it 
has been argued that neutrinos arising from pion decay in GRB fireballs
may be detectable. This neutrino signal depends, along with the 
other parameters characterizing the fireball, on the electron fraction in 
the fireball. If more realistic GRB central engine models also 
leave their finger prints in the electron fraction in the outflow,
 such a signal
might be used to distinguish between central engine models.  

\acknowledgements 

This work was partially supported by NSF Grant PHY98-00980 at UCSD, and
an  IGPP mini-grant at UCSD. We are indebted to C.Y. Cardall for the 
use of an unpublished paper on the relativistic formulation of the 
wind equations.

\begin{figure*}
\epsscale{1.}
\plotone{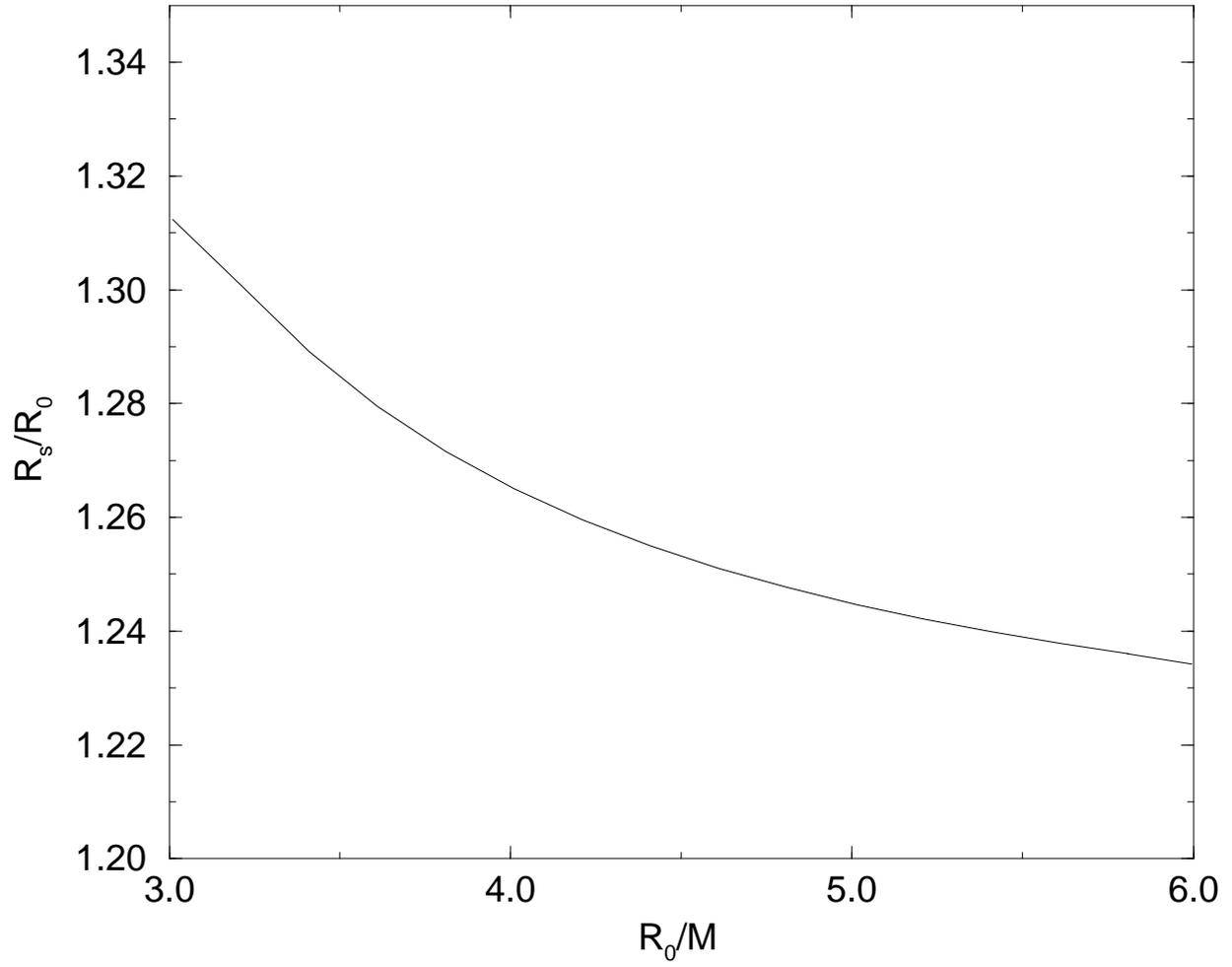}
\caption{\small Approximate position of the the sonic point as a function of $r_0/M$ for 
neutrino anti-neutrino annihilation dominating the heating term within 
the sonic radius. The momentum depostion term $a$ has been neglected and it 
is assumed that at the sonic point $H(1-g)\gg 1$.
}
\end{figure*}

\begin{figure*}
\epsscale{1.}
\plotone{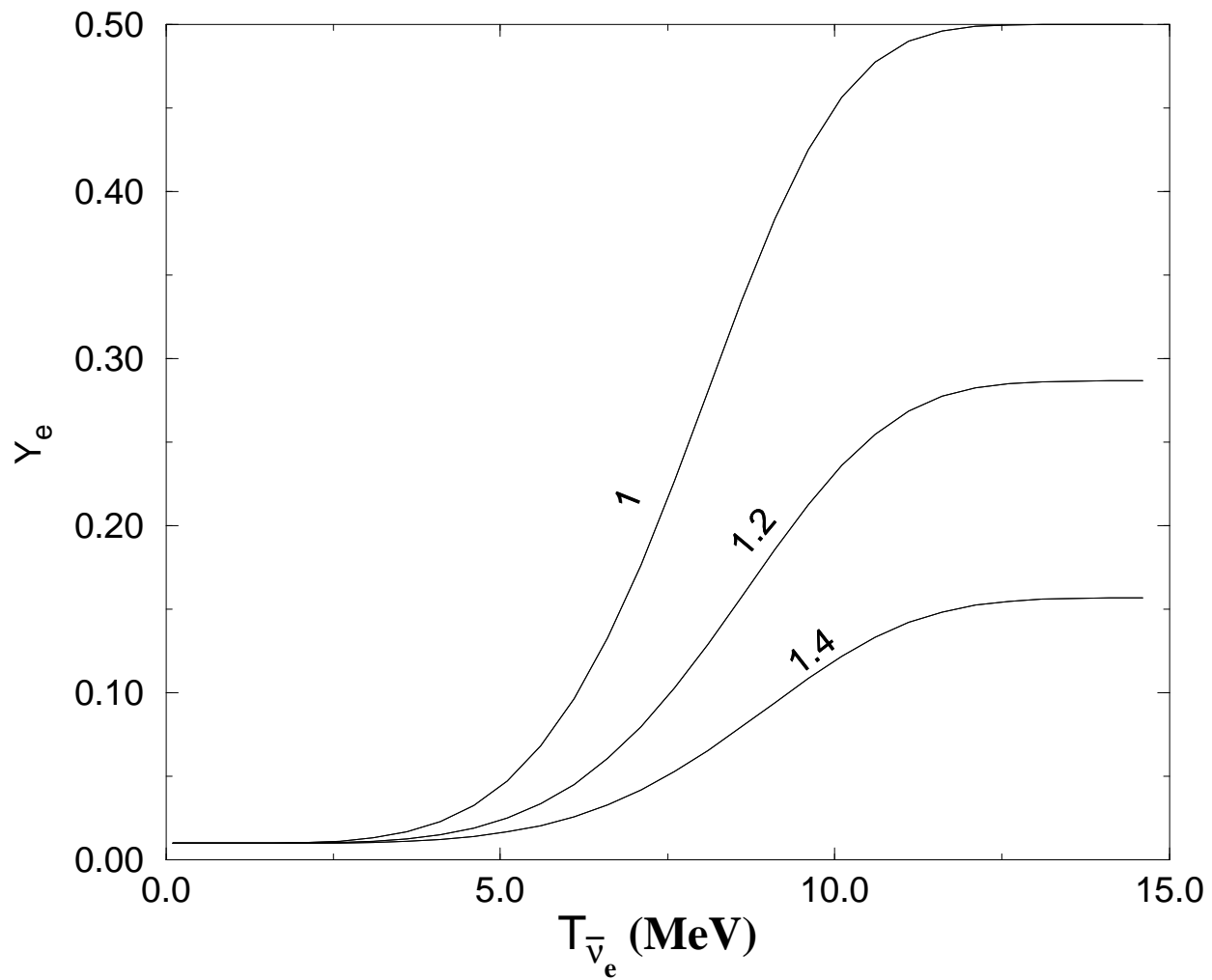}
\caption{\small Final electron fraction in the fireball as a function of 
$T_{\bar \nu_e}$ and for different ratios  $T_{\bar \nu_e}/T_{\nu_e}$ (labeled
next to the curves). The initial electron fraction is taken to be 0.01, and 
the entropy of the plasma is assumed to increase by a factor of $10^5$ as 
the plasma travels from the neutron star surface to the sonic point.
}
\end{figure*}

\begin{figure*}
\epsscale{1.}
\plotone{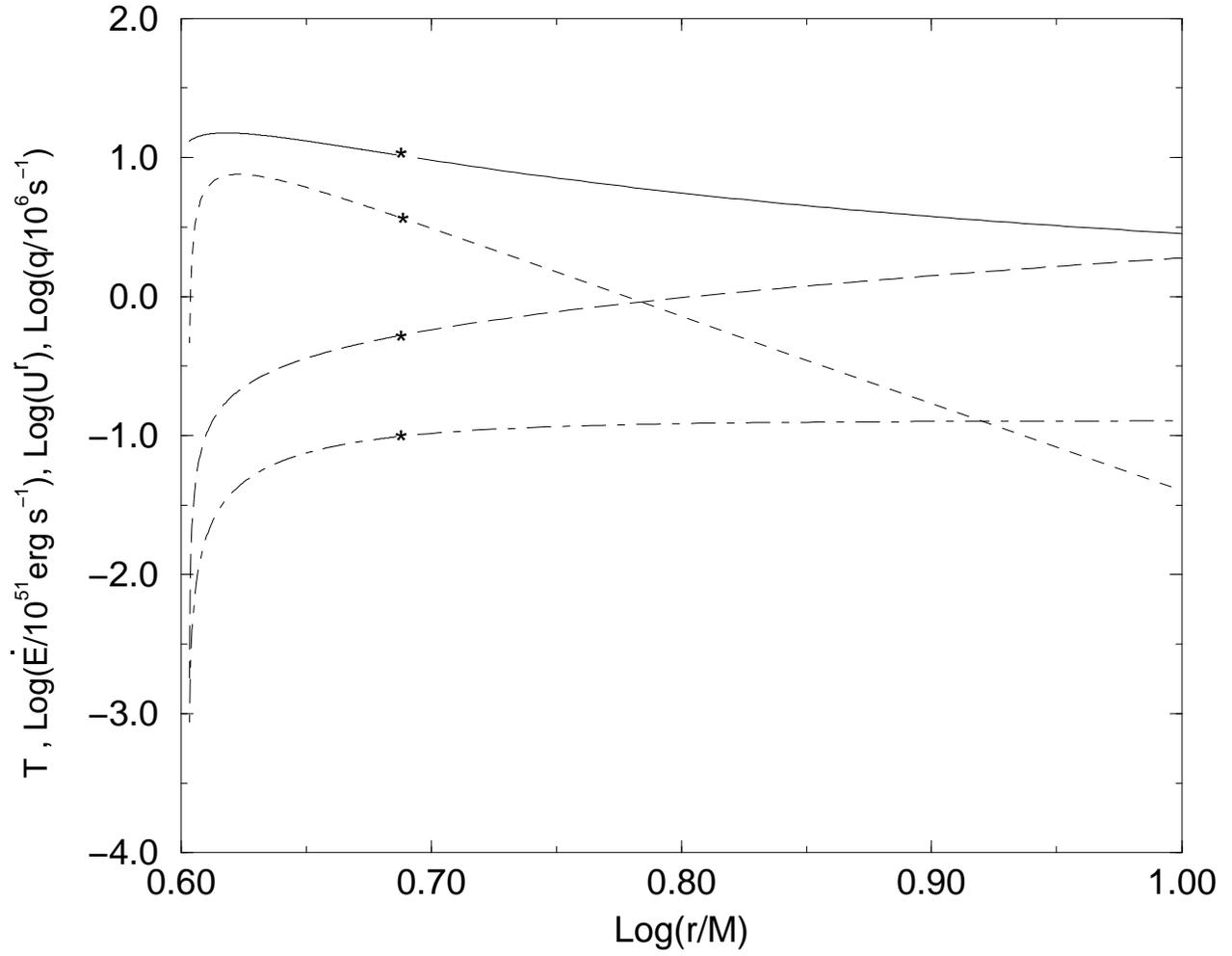}
\caption{\small 
Evolution of a neutrino heated wind solution as described in the text.
 The solid curve is for 
temperature (in MeV), the short dashed curve for $q_{\nu \bar{\nu}}$,
the long dashed curve for $U^r$, and the dot-dashed curve for
$\dot E=\dot M \hat H y$. The asterices correspond to 
the position of the sonic point.}
\end{figure*}

\end{document}